%% file: sample-acmtog.tex
\documentclass[sigconf]{acmart}
\AtBeginDocument{%
  }

\setcopyright{acmlicensed}
\copyrightyear{2018}
\acmYear{2018}
\acmDOI{XXXXXXX.XXXXXXX}

\begin{document}

\title{Efficient Large-Scale Cross-Domain Sequential Recommendation with Dynamic State Representations}


\author{Manuel V. Loureiro}
\email{manuel.loureiro@huawei.com}
\affiliation{%
  \institution{Huawei Ireland Research Centre}
  \city{Dublin}
  \country{Ireland}
}
\author{Steven Derby}
\email{steven.derby@huawei-partners.com}
\affiliation{%
  \institution{Huawei Ireland Research Centre}
  \city{Dublin}
  \country{Ireland}
}

\author{Aleksei Medvedev}
\email{aleksei.medvedev@huawei-partners.com}
\affiliation{%
  \institution{Huawei Ireland Research Centre}
  \city{Dublin}
  \country{Ireland}
}
\author{Alejandro Ariza-Casabona }
\email{alejandro.ariza.casabona@h-partners.com}
\affiliation{%
  \institution{Huawei Ireland Research Centre}
  \city{Dublin}
  \country{Ireland}
}
\author{Gonzalo Fiz Pontiveros}
\email{gonzalo.fiz.pontiveros@huawei.com}
\affiliation{%
  \institution{Huawei Ireland Research Centre}
  \city{Dublin}
  \country{Ireland}
}
\author{Tri Kurniawan Wijaya}
\email{tri.kurniawan.wijaya@huawei.com}
\affiliation{%
  \institution{Huawei Ireland Research Centre}
  \city{Dublin}
  \country{Ireland}
}
\authornote{Manuel V. Loureiro and Steven Derby contributed equally to this work.} 

\keywords{Auto-regressive models,
multi-domain recommendation,
sequential recommendation,
transformers,
retrieval.}

\received{20 February 2007}
\received[revised]{12 March 2009}
\received[accepted]{5 June 2009}


\input{00_abstract}
\maketitle 

\input{01_introduction}
\input{02_related_work}

\input{03_methodology}

\input{04_experiments}
\input{05_conclusion}

\bibliographystyle{ACM-Reference-Format}
\bibliography{references}

\appendix









\end{document}

%% file: 00_abstract.tex
\begin{abstract}
Recently, autoregressive recommendation models (ARMs), such as Meta's HSTU model, have emerged as a major breakthrough over traditional Deep Learning Recommendation Models (DLRMs), exhibiting the highly sought-after scaling law behaviour. However, when applied to multi-domain scenarios, the transformer architecture's attention maps become a computational bottleneck, as they attend to all items across every domain.
To tackle this challenge, systems must efficiently balance inter and intra-domain knowledge transfer. In this work, we introduce a novel approach for scalable multi-domain recommendation systems by replacing full inter-domain attention with two innovative mechanisms: 1) \textbf{Transition-Aware Positional Embeddings (TAPE):} We propose novel positional embeddings that account for domain-transition specific information. This allows attention to be focused solely on intra-domain items, effectively reducing the unnecessary computational cost associated with attending to irrelevant domains. 2) \textbf{Dynamic Domain State Representation (DDSR):} We introduce a dynamic state representation for each domain, which is stored and accessed during subsequent token predictions. This enables the efficient transfer of relevant domain information without relying on full attention maps. Our method offers a scalable solution to the challenges posed by large-scale, multi-domain recommendation systems and demonstrates significant improvements in retrieval tasks by separately modelling and combining inter- and intra-domain representations.
\end{abstract}

%% file: 01_introduction.tex
\section{Introduction}
Cross-domain sequential Recommendation Systems (CDSR) have emerged as a key development over the more traditional Sequential Recommendation (SR) approaches for businesses. In this setting, the system is tasked with providing recommendations for users based on past historical interactions across multiple domains. By extending the system to a cross-domain strategy, it is possible to employ cross-domain information; different domains may contain complementary knowledge about user behaviour. For instance, a recommendation system aiming to suggest a new film to a particular user would benefit from information based on the user's history of purchased books. Cross-domain approaches can improve recommendations by leveraging interactions across multiple domains. However, treating these interactions as a single history is computationally demanding, in particular as transformer networks scale quadratically with sequence length. Alternatively, splitting and processing sequences independently by domain may fail to capture cross-domain similarities.
Cross-domain systems need to effectively balance both inter- and intra-domain knowledge transfer to overcome the challenges associated with cross-domain recommendation.

In this work, we focus on transformer architectures and explore strategies in which self-attention operations are restricted to intra-domain processing, for which there is limited current work. Instead of attention-based mechanisms, we capture inter-domain knowledge via two novel approaches: 1) Transition-Aware Positional Embeddings (TAPE), which capture cross-domain transitions in the input before intra-domain processing, and 2) Dynamic Domain State Representation (DDRS), which accumulates the latest domain state representation at each position and uses it to capture deeper cross-domain insights through attention. Through theoretical and experimental analysis, we demonstrate that these methods reduce computational cost whilst preserving strong overall performance on next-item prediction tasks. Further findings from ablation experiments verify the effectiveness of our proposed approach.

%% file: 02_related_work.tex
\section{Related Work}
Sequential models like the \textbf{GRU} \cite{hidasi2015session} were originally introduced to efficiently handle long sequences in recommendation-based sessions. However, recent advancements have seen transformer-based models such as \textbf{Bert4Rec} \cite{sun2019bert4rec}, \textbf{SASRec} \cite{kang2018self}, and \textbf{HSTU} \cite{zhai2024actions}, displaying superior performance and scalability. Our contributions are built on top of the latter for its mentioned capabilities that make it suitable for a large-scale industrial application.

More recently, state-based models, like \textbf{Mamba} \cite{liu2024mamba4rec}, \textbf{ECHO} \cite{wang2024echomamba4rec}, and \textbf{Sigma} \cite{liu2025sigma}, are being explored as a promising direction. Among these sequential recommendation models, cross-domain sequential recommendation has surfaced as a significant research area. \textbf{$\pi$-Net} \cite{ma2019pi} and \textbf{Parallel Split-Join} \cite{sun2021parallel} enable domain transfer using a cross-domain state that updates upon domain shifts, along with shared-account filtering. The model \textbf{MAN} \cite{lin2024mixed} captures both single-domain and cross-domain knowledge using mixed attention on item sequences at local and global scales. Gao et al. present \textbf{C2DSR} \cite{cao2022contrastive}, which leverages item-item interactions to build graphs for single- and cross-domain inputs, applying attention encoders for next-item prediction. Instead, Xiao et al. \shortcite{xiao2023proxy} create proxy-aware item representations utilizing textual features and time-interval-aware attention encoders, enabling cross-domain sequential recommendation through contrastive learning. \textbf{DREAM} \cite{ye2023dream} also employs contrastive learning and enhances domain knowledge transfer through domain extraction and attention mechanisms across single and cross domains. Whilst pragmatic, such approaches are often ill-suited to the parallel computations of transformers or their scaling with model complexity. 
Thus, we present a novel technique to effectively capture dynamic domain transitions using transformers with intra-domain partition computation.

%% file: 03_methodology.tex
\section{Methodology}
In this section, we outline the setting in which we operate, as well as the background and theoretical motivation, before presenting the novel techniques we employ to achieve our objectives. A particular distinction between state-based models and transformer networks in CDSR lies in how multi-domain sequences are handled. In state-based models, it is easy to traverse contiguous sequences while performing discrete procedures on particular positions of interest. For instance, caching a domain-state representation and updating it as transitions occur in the sequence has been done to good effect \cite{ma2019pi,sun2021parallel}. On the other hand, transformers perform computations in parallel for each time step, making such dynamic adjustments trickier. Some work computes individual intra-domain sequence representations to capture each domain yet still requires multiple passes \cite{ye2023dream}.
We instead focus on an approach that treats the history as a single sequence --- maintaining parallel computations --- while computing only intra-domain attention; that is, attention is computed only within subsequences that share the same domain which is both computationally efficient and effective for CDSR. 

\subsection{Problem Statement}
Let $\mathcal{I}$ denote the set of items and assume there exist a partition $\mathcal{I}=\mathcal{I}_1\sqcup\ldots\sqcup\mathcal{I}_D$ into $D$ domains. We consider a dataset $\mathcal{U}$ consisting of tuples $(u,S_u)$ , where $S_{u} = \{x_{1}, x_{2}, \dots, x_{n} \}$  is the interaction history of user $u$. Furthermore, let $T_{u} = \{d_{1}, d_{2}, \dots, d_{n} \}$ be the corresponding domain sequence; that is, $x_{i} \in \mathcal{I}_{d_i}$ for $1\leq i \leq n$.

We define our CDSR objective as learning a parametrised function $F: \mathcal{I}^{n} \times \ \mathcal{D}^{n} \times \mathcal{D} \rightarrow \left[ 0,1 \right]$ that estimates the probability of the next item of a sequence $x \in \mathcal{I}$ given its domain $d \in \mathcal{D}$ i.e.  $F(S_u, T_u, d)~\sim~P(x| S_u, T_{u}, d)$. Note that, in contrast to contemporary work where the domain is also inferred by the model, we endow the model with domain information as a prior, since such information is readily available in commercial scenarios.
The function $F$ is implemented as a transformer network which takes as input a set of embeddings in sequential order to predict the next token. 

\subsection{Transition-Aware Positional Embeddings (TAPE)}
We define the embedding table such that each input sequence at position $i$ consists of embeddings $e_{i} \in E \in \mathbb{R}^{|\mathcal{I}| \times k}$, $k \in \mathbb{N}$. Our goal here is to define positional embeddings that also convey when there is a domain transition: a point where the next item belongs to a different domain. To achieve this, we introduce domain embeddings $\hat d$ for all $d \in \mathcal{D}$. For input sequence embeddings $\{e_{1}, e_{2}, \dots, e_{n}\}$ with domain embeddings $\{\hat d_{1}, \hat d_{2}, \dots, \hat d_{n}\}$, we update the embedding $e_{i} \in \mathbb{R}^{k}$ belonging to domain $d_{i}$ as follows:
\begin{align}
    \hat{e_{i}} &= e_{i} + p_{i} + r_{i}, \label{eq:1}\\
    r_{i} &= \begin{cases} 
       \hat d_{i+1}\odot(W\cdot \hat d_{i} + b) & d_{i+1} \neq  d_{i} \\
      0 &  d_{i+1} = d_{i}
   \end{cases},
\end{align}
\noindent where $\hat d_i = M(d_{i})$ for some learnable $M \in \mathbb{R}^{|\mathcal{D}|\times k}$ and $W \in \mathbb{R}^{k\times k}, b \in \mathbb{R}^k$ are learnable parameters. Finally, $p_{i}$ denotes the usual absolute positional embedding in transformers. TAPE includes a signal about when and how a shift occurs, so the model can adapt to the new domain space without incurring large computational overhead.
\subsection{Sequential Masking Layers}

In the CDSR literature, there are generally two methods by which sequences are processed. The first relies on separation: individual intra-domain sequences are processed independently and rely on specific mechanisms for knowledge transfer. The second uses complete end-to-end inter-domain sequence processing and all domains are treated in a single sequence with some underlying approach to capture domain specificity.
Each approach has benefits and drawbacks, and some models even combine both methods to their advantage. The former is more efficient but relies heavily on the model to capture these information streams. The latter tends to perform well but is computationally intensive --- especially in transformer-based architectures, where sequences scale quadratically.

In this work, we explore the former when applied to transformer networks, processing sequences to create theoretical savings in computation. We begin by defining our standard query, key, and value matrices $Q^{L}, K^{L}$, and $V^{L}$, respectively, along with one extra matrix $U^{L}$. We generate each matrix by applying pointwise projections along the sequence dimensions of the previous input layer $H^{L-1} \in \mathbb{R}^{n \times k}$ and split into four distinct matrices:


\begin{equation}
    Q^{L}, K^{L}, V^{L}, U^{L} = \text{Split}(W_{a} \cdot H^{L-1} + b_{a}) \label{eq:2}
\end{equation}

\noindent with $W_{a} \in \mathbb{R}^{k \times 4k}$ and $b_{a} \in \mathbb{R}^{4k}$. They are then separated into attention heads $h \in \mathbb{N}$ such that $Q^{L}, K^{L}, V^{L}, U^{L} \in \mathbb{R}^{h \times n \times k/h}$.
We define our self-attention layer as follows. For a particular layer $L$:

\begin{equation}
    A^L = \text{Softmax}\left( \dfrac{FlexAtt(Q^{L}, K^{L}, T_u ) + \text{ALiBi}}{\sqrt{k}}\right) \label{eq:3}
\end{equation}


\noindent where \textbf{ALiBi} is the positional encoding \cite{press2021train}, and \textit{FlexAtt} denotes the flex attention operation \cite{li2024flexattention} which we use to facilitate computations of only intra-domain representations. Concretely, given the positions $i,j$, when $ d_i = d_j$ the value of the key-query calculation between the $i$-th and $j$-th elements of the sequence is computed as $Q_{i}^{L} \cdot (K_{j}^{L})^{T}$ and skipped otherwise. This approach avoids unnecessary computations for computing intra-domain representations at each position in the sequence.

Our final output is then given by the standard feed-forward operation in self-attention with our $U^{L}$, which acts as a gating weight following \textbf{HSTU} \cite{zhai2024actions}:


\begin{equation}
    H^{L} = \text{FFN}\left(\text{Norm}(A^{L}, V^{L}) \odot U^{L})\right) \label{eq:4} 
\end{equation}

\subsection{Dynamic Domain State Representation}

We propose a modification over existing domain state representations that incorporates the most recent hidden state of each domain, which is more suitable to our case where only intra-domain partitions are computed. We define our domain state matrix $H_{\mathcal{D}}^{L} \in \mathbb{R}^{|\mathcal{D}| \times n \times k}$ which consists of the last hidden state for a particular domain at each step. Specifically, for each domain $d \in \mathcal{D}$ at position $1 \leq i \leq n$, we select the last hidden state up to $i$ that belongs to the domain $d$ using a selection function $\phi_{d}(i) = max\{j: d = T_{u, j} , \, 1 \leq j \leq i \}$. Letting $(H_{\mathcal{D}}^{L})^{(d, i)}$ define the hidden state representation of $H_{\mathcal{D}}^{L}$ at position $i$ given domain row $d$, we define our matrix as $(H_{\mathcal{D}}^{L})^{(d, i)} = H_{\phi_{d}(i)}^{L-1}$ if the domain $d$ has occurred at some point or at $i$ (that is, $\{j: d = T_{u, j} , \, 1 \leq j \leq i\} \neq \emptyset$), otherwise its a zero vector. Finally, we define our dynamic domain state representation module as:



\begin{equation}
    K^{L}, V^{L} = \text{Split}(W_{c} \cdot H^{L-1} + b_{c}), \quad W_c \in \mathbb{R}^{k \times 2k}, b_c\in \mathbb{R}^{2k} \label{eq:5}
\end{equation}

\begin{equation}
    Q^{L} = W_q \cdot H_{\mathcal{D}}^L + b_q, \quad W_q \in \mathbb{R}^{k \times k}, b_c\in \mathbb{R}^{k}
\end{equation}

\begin{equation}
     C^{L} = \text{Norm} \left(  \sum_{d \in D}\text{Softmax}\left(\dfrac{ Q^{L} \cdot K^{L}}{\sqrt{k}}\right) \cdot V^{L} \right)
\end{equation}

This formulation is relatively computationally cheap as typically $n>>|\mathcal{D}|$,  yet effective for knowledge transfer in multi-domain sequential recommendation\footnote{If accepted, we will provide more details on the Flex Attention block masking used in this work and also how to generate this domain state representation.}. We apply this module as an extra self-attention layer, following a similar pattern to equations \ref{eq:2} and \ref{eq:3}. Finally, at each layer, we combine outputs as $\hat{H}^{L} = H^{L} + C^{L}$.

\subsection{Complexity}
Let $S = \sum_{d \in \mathcal{D}} s_d$ be the size of our sequences where $s_d$ denotes the number of items in domain $d$  and let $D=|\mathcal{D}|$. The complexity of  our self-attention mechanism is $O\left (\sum_{d \in \mathcal{D}} s_d^2\right )$ whereas classical attention scales like $S^2=\left (\sum_{d \in \mathcal{D}} s_d \right)^2 $ . Optimality occurs when $s_d = \frac{S}{D}$ for all $d \in \mathcal{D}$, where the complexity is $O(\frac{1}{D}|S|^2)$. In general, however, we can only say that the complexity is bounded by $O(\delta S^2)$ where $\delta = \max_{d}\frac{s_d}{S}$. \footnote{The max should also be taken over all possible sequences.}



%% file: 04_experiments.tex
\begin{table*}[t]
\centering
\scriptsize
\begin{tabular}{l|l|l|l|l|l|l|l|l|l|l|l}
\toprule
    \textbf{Model} & \textbf{HR@1} & \textbf{HR@10} & \textbf{HR@50} & \textbf{HR@100} & \textbf{HR@200} & \textbf{MRR} & \textbf{NDCG@10} & \textbf{NDCG@50} & \textbf{NDCG@100} & \textbf{NDCG@200} \\
\midrule \specialrule{.1em}{.05em}{.05em}
HSTU (w/ RAB) & 0.57 (0.12) & 2.04 (0.46) & 4.38 (1.01) & 5.90 (1.36) & 7.84 (1.78) & 1.14 (0.23) & 1.21 (0.26) & 1.71 (0.37) & 1.96 (0.42) & 2.23 (0.48) \\
HSTU (w/ ALiBi) & 0.66 (0.07) & 2.11 (0.09) & 4.34 (0.09) & 5.79 (0.16) & 7.61 (0.26) & 1.22 (0.08)  & 1.29 (0.09) & 1.77 (0.07) & 2.01 (0.06) & 2.26 (0.06) \\
Intra-domain Masking (w/ALiBi) & \textbf{0.99} (0.05) & 3.36 (0.03) & 6.86 (0.25) & 9.09 (0.41) & 11.90 (0.57) & \textbf{1.88} (0.03) & \textbf{2.03} (0.03) & 2.78 (0.02) & 3.15 (0.05) & 3.54 (0.07) \\
\textbf{Ours} & 0.80 (0.02) & \textbf{3.40} (0.03) & \textbf{7.59} (0.06) & \textbf{10.25} (0.07) & \textbf{13.55} (0.06) & 1.77 (0.02) & 1.92 (0.02) &
\textbf{2.82} (0.03) & \textbf{3.26} (0.03) & \textbf{3.72} (0.03) \\
\midrule
Ablation (w/o DDSR) & 1.00 & 3.36 & 6.82 & 9.01 & 11.77 & 1.89 & 2.04 & 2.79 & 3.14 & 3.53 \\
Ablation (w/o TAPE) & 0.76 & 3.20 & 7.23 & 9.82 & 13.09 & 1.67 & 1.80 & 2.67 & 3.09 & 3.55 \\
\bottomrule
\end{tabular}
\caption{Evaluation results in percentage points. HR = Hit Rate, MRR = Mean Reciprocal Rank, NDCG = Normalized Discounted Cumulative Gain. Except for the ablations, all values are the mean over 3 runs, with standard deviation in parentheses. "Ours" denotes the full model with intra-domain masking, TAPE, and DDSR.}
\label{results}
\end{table*}

\section{Experiments}
Following the training objective used in \textbf{HSTU} \cite{zhai2024actions}, we perform causal autoregressive prediction of the next item in the sequence. Accordingly, we apply a causal mask in the self-attention layers and compute the next-item probability at every step of the sequence, similar to \textbf{HSTU} \cite{zhai2024actions} and \textbf{SASRec} \cite{kang2018self}, limiting the probability space to only those items within the same domain.
Concretely, we use the item embeddings to compute logits for each item based on its domain. This is achieved by restricting the embedding space to the items belonging to the domain of the next item. Given the final output of the self-attention blocks $H^{L}$, we compute the probability for each item representation $h_{i} \in H^{L}$ as:

\begin{equation}
    P(x_{i+1} \mid d_{i+1}) = \text{SampledSoftmax}\left(h_{i}\left(E^{d_{i+1}}\right)^{\mathsf{T}}\right)
\end{equation}

\noindent where `SampledSoftmax` \cite{covington2016deep} computes softmax probabilities over a reduced space by sampling a number of negative examples. Since we constrain the next-item prediction to a specific domain, negative samples are also selected only from that domain. The final loss is computed as the negative log-likelihood of the predicted next item.

\subsection{Dataset}
To empirically evaluate the performance of our approach, we utilize the Amazon cross-domain datasets \cite{mcauley2015image}, which include numerous user-item interactions across multiple domain categories. To demonstrate the utility of our approach, we incorporate five domain categories in our dataset: \textit{Books}, \textit{Movies and Tv}, \textit{Electronics}, \textit{Cds and Vinyls}, and \textit{Clothing, Shoes and Jewelry}.
Unlike previous work that typically focuses on only two domains and struggles to scale with the addition of more, we aim to demonstrate how our model scales both in terms of operational cost and downstream performance.
To preprocess the data, we combine all domain datasets and remove users and items that occur fewer than 5 times. We then combine user histories, containing items and domains, by ordering by timestamp. Finally, we filter out any sequence with fewer than 10 interactions.

\subsection{Evaluation, Model Performance and Ablation Study}
Following prior work, we adopt the \textit{leave-one-out} strategy for training and evaluation. For each sequence $S^{\mathcal{X}}_{u} \in \mathcal{U}$, we train the model autoregressively on the first $|S^{\mathcal{X}}_{u}| - 1$ items. During evaluation, the entire sequence $S^{\mathcal{X}}_{u}$ is provided, and the final item, unseen during training, is used for performance measurement. Metrics are reported on this final prediction using the standard ranking-based measures Hit Rate (HR), Mean Reciprocal Rank (MRR), and Normalized Discounted Cumulative Gains (NDCG). HR and NDCG are evaluated at various $k$ values.

The baseline for our experiments is the original HSTU model, which uses full self-attention over user sequences along with a learned relative attention bias (RAB). As a first variant, we replace RAB with ALiBi \citep{press2021train}, a static, non-learned bias that scales linearly with distance. This leads to small but consistent gains in NDCG while HR improves only for smaller top-$k$ values.

A more substantial change comes from enforcing intra-domain attention masking using FlexAttention \cite{li2024flexattention}. Instead of allowing all tokens to attend to one another, we restrict the model to operate within each domain boundary independently and limit the output vocabulary only to the domains of the next item. This intra-domain isolation significantly improves performance across the board, especially at higher $k$ thresholds). These gains suggest that in multi-domain settings, forcing the model to focus on a single domain at a time might avoid the noise introduced by modelling inter-domain dependencies.

Our full model builds upon this intra-domain setup by incorporating TAPE, which explicitly encodes domain transitions, and DDSR, which propagates high-level representations across domains. When both are included, the model demonstrates consistent improvements across all evaluation metrics. Compared to the masked-only configuration, Hit Rate improves from 9.09 to 10.25 at HR@100, and NDCG@100 rises from 3.15 to 3.26. At the same time, lower $k$ metrics like HR@1 and MRR see smaller drops (from 0.99 to 0.80, and from 1.88 to 1.77 respectively), suggesting a slight tradeoff in ranking sharpness but with stronger overall recall. Importantly, this configuration maintains high precision at larger retrieval depths, making it well-suited for generating candidate pools for downstream ranking tasks.

To analyse the individual contribution of our novel mechanisms, we retrain the models with and without each component over a single run. Results are also presented in Table \ref{results}. 

\subsubsection{w/o DDSR}
Including only TAPE in our multi-domain model demonstrates strong performance in the lower @K values but suffers at higher @K values Notice that while TAPE's performance is similar to intra-domain masking, only including DDSR as above significantly drops our model's performance. We believe this may be due to TAPE's ability to facilitate domain transfer, which further demonstrates the overall necessity for their inclusion.

\subsubsection{w/o TAPE}
On the other hand, removing TAPE seems to have the opposite effect in that the model performs worse at lower values of @K but performs best for larger values of K indicating that DDSR leads to a broader, more diverse, distribution over next items. \footnote{Since this is early work, our evaluation is limited to our baseline model. In future work, we plan to compare with other state-of-the-art approaches.}

\subsection{Discussion}
Removing TAPE or DDSR hurts performance, but differently: TAPE mainly affects ranking (NDCG@100 ↓ 3.14), while DDSR impacts recall (HR@100 ↓ 9.82). This suggests a functional split—DDSR supports exploration by transferring cross-domain context; TAPE focuses the model on domain-specific exploitation. The intra-domain masking baseline, which lacks TAPE, shows similar issues, indicating that isolating domains reduces prediction diversity and confirming DDSR’s role. This points to a structural limitation in parallel intra-domain models. Domain separation may also amplify popularity bias \cite{klimashevskaia2024survey}, though we leave this to future work.

%% file: 05_conclusion.tex
\section{Conclusion}

We explore how transformer models can process intra-domain sequences in parallel and adapt this setup for CDSR. To enable cross-domain transfer without significant overhead, we introduce two mechanisms: TAPE, which signals domain transitions, and DDSR, which incorporates cross-domain context into intra-domain processing. Our method improves over the baseline and highlights the promise of this formulation for CDSR. While our current implementation is not yet optimized, theoretical analysis indicates clear runtime benefits, which we aim to confirm in future work. These results are preliminary but point to a viable and efficient path forward for cross-domain recommendation.